\documentstyle[psfig,epsfig,aps,aps12]{revtex}
\def\be{\begin{equation}}
\def\ee{\end{equation}}
\def\bea{\begin{eqnarray}}
\def\eea{\end{eqnarray}}
\def\lbl{\label}

\begin{document}
\draft
\title{Perturbations of brane worlds}
\author{Nathalie Deruelle$^{1,2,3}$, Tom\'a\v s Dole\v zel$^{1,4}$ and Joseph Katz$^{5,6}$}
 \address {$^1$ D\'epartement d'Astrophysique Relativiste et de Cosmologie,\\ 
UMR 8629 du Centre National de la Recherche Scientifique,\\ 
Observatoire de Paris, 92195 Meudon, France}
 \address{$^2$ Institut des Hautes Etudes Scientifiques,\\ 
91140 Bures-sur-Yvette, France}
 \address{$^3$ Centre for Mathematical Sciences, DAMTP,\\ 
University of Cambridge, Wilberforce Road, Cambridge, CB3 0WA, England}
 \address{$^4$ Institute of Theoretical Physics, Charles University,\\ 
V Hole\v sovi\v ck\'ach 2, 18000 Prague 8, Czech Republic}
 \address{$^5$ Racah Institute of Physics, Hebrew University,\\
91904, Jerusalem, Israel}
 \address{$^6$ Institute of Astronomy, University of Cambridge,\\
Madingley Road, Cambridge, CB3 0HA, England}

\date{\today}

\maketitle

\begin{abstract}
We consider cosmological models where the universe, governed by Einstein's equations, is a piece of a five dimensional
double-sided anti-de Sitter spacetime (that is, a ``$Z_2$-symmetric bulk") with matter confined to its four 
dimensional Robertson-Walker boundary or ``brane". We study the perturbations of such models.  We use conformally
minkowskian coordinates to disentangle the contributions of the bulk gravitons and of the motion of the brane. We find
the restrictions put on the bulk gravitons when matter on the brane is taken to be a scalar field and we solve in this
case the brane perturbation equations.
\end{abstract}


%------------ Paper -----------------------------------------------------------------------------------------------------
\section*{I Introduction}
In a now classic paper [1] Randall and Sundrum indicated how one could recover the linearized Einstein equations on a
four dimensional minkowskian ``brane", a brane being in that context a boundary of a five dimensional double-sided
anti-de Sitter spacetime ($AdS_5$), that is of a ``$Z_2$-symmetric bulk". This discovery was soon followed by the building of
cosmological models, where the brane, instead of flat, is taken to be a Robertson-Walker spacetime, and it was shown
that such ``brane worlds" can tend at late times to the standard big bang model and hence represent the observed
universe (see e.g. [2] for early models and [3-5] for fully relativistic ones).

More recently various theoretical set ups to study the perturbations of such cosmological models have been proposed
[6-13]. The purpose of these analyses is, in particular, to eventually calculate the Cosmic Microwave Background
anisotropies predicted by brane worlds. However, they all have up to now stalled on the problem of solving, in a general
manner,  the Lanczos-Darmois-Israel equations (that is the Einstein equations integrated across the brane, often called
``junction conditions") which relate the matter perturbations on the brane and the perturbations in the bulk.

In order to be in a position to solve these equations, we present in this paper the perturbation theory of brane worlds
from a geometrical point of view, in the line of [6], [7] and [10]. This approach, which uses conformally minkowskian
coordinates that are well adapted to the geometry of the bulk, will first allow a clear and simple distinction between
the perturbations in the brane due to perturbations in the bulk and the perturbations in the brane due to its motion. (As
we shall comment upon,  the latter, so-called ``brane-bending" effect, is more difficult to describe when gaussian
coordinates are used, as in e.g. [8], [9], [11] or [13].)

We shall then write the Lanczos-Darmois-Israel equations and see that only a sub-class of bulk gravitational waves is
compatible with a given type of matter on the brane. As an example we shall consider the case when matter on the brane
is imposed to be a scalar field and find explicitly in that case the allowed bulk gravitational waves. We shall then solve
the Lanczos-Darmois-Israel equations and give in a  closed form the perturbed metric and scalar field in the brane.

The paper is organized as follows. In section 2 we present the formalism and notations used and describe the background
brane and bulk. Section 3 treats the geometry and matter perturbations in the brane induced by a bending of the brane in
a strictly anti-de Sitter bulk. As for section 4, it considers the changes in the brane induced by perturbations of the bulk.
There is nothing essentially new in these sections 3 and 4, but the presentation is, we hope, more straightforward and
pedagogical than some. In section 5 we dwell on gauge issues, count the degrees of freedom of the perturbations in the
brane and comment upon the use of gaussian normal coordinates in which, as we shall argue, the brane bending effect is
described in a fairly awkward manner. We recall in section 6 standard results of the linearized Einstein equations
in an anti-de Sitter spacetime in conformally minkowskian coordinates. In section 7 the Lanczos-Darmois-Israel
equations are written and solved when matter on the brane is taken to be a scalar field.

\section*{II The background bulk and brane in conformally min\-kowskian coordinates}
The ``bulk'' is a piece of a five dimensional spacetime of which the  four dimensional edge, or ``brane",  is supposed to
represent our universe. At zeroth order in perturbation theory this background bulk will be chosen to be an anti-de Sitter
spacetime (see below a reason why we do not consider a de Sitter bulk and e.g. [14-15] and references therein for more
general backgrounds).

Many different coordinate systems can be used to describe anti-de Sitter spacetime: see e.g. [1] or [5] for normal
gaussian coordinates in which the surface $y=0$ represents the brane, e.g. [16] for Schwarzschild-like coordinates 
and [17] for their local equivalence. In this paper we shall use conformally minkowskian coordinates $X^A$, in which the
metric of  a five dimensional anti-de Sitter spacetime reads
\be d\bar s^2|_5=\bar g_{AB}\,dX^AdX^B\quad\hbox{with}\quad \bar g_{AB}={1\over ({\cal K}X^4)^2}\,\eta_{AB}\lbl{(2.1)}\ee
where the upper case indices $A,B$ run from $0$ to $4$, $\eta_{AB}={\rm diag}(-1,1,1,1,1)$ and ${\cal K}$
is a positive constant. Note that the coordinates $X^A$ do not cover the whole $AdS_5$ spacetime; see appendix A for details.

The background brane is a four dimensional surface in $AdS_5$ with maximally symmetric spatial sections. We shall
restrict our attention to those Robertson-Walker branes which have euclidean spatial sections. The equation for such a
brane
$\bar\Sigma$ is
\be X^A=\bar X^A(x^\mu)\qquad\hbox{with}\qquad \bar X^0=T(\eta)\quad,\quad \bar X^i=x^i\quad,\quad
\bar X^4=A(\eta)\lbl{(2.2)}\ee 
where the four coordinates $x^\mu \,(x^0\equiv\eta, x^i)$, lower case latin
indices running from 1 to 3,  parametrize the brane, where
$A(\eta)$ is an a priori arbitrary function of  $\eta$ and where $T(\eta)$ is defined up to
an arbitrary constant by
\be T'=\sqrt{1+A'^2}\lbl{(2.3)}\ee
a prime denoting a derivative with respect to $\eta$. This condition defines $\eta$ as conformal time.
Four independent tangent vectors to
the brane are
\be \bar V^A_\mu\equiv {\partial \bar X^A\over\partial x^\mu}\qquad\hbox{that is}\qquad \bar V^A_\eta=(T',0,0,0,A')
\quad\hbox{,}\quad \bar V^A_i=(0,\delta^A_i,0).\lbl{(2.4)}\ee
The induced metric on the brane is also conformally minkowskian and reads
\be d\bar s^2|_4=\bar g_{AB}|_{\bar\Sigma}\,\bar V^A_\mu\bar V^B_\nu\,dx^\mu dx^\nu  =
{1\over ({\cal K}A)^2}\,\eta_{\mu\nu}\,dx^\mu dx^\nu .\lbl{(2.5)}\ee 
It will be useful in the following to introduce the scale factor $a(\eta)$, the cosmic time $t$,  and
the Hubble parameter
$H$ defined by
\be a\equiv{1\over {\cal K}A}\quad,\quad dt\equiv ad\eta\quad,\quad H\equiv{\dot a\over a}\lbl{(2.6)}\ee
where a dot denotes a derivative with respect to $t$.

At this stage one can note in passing that if the bulk had been chosen to be a de Sitter rather than an anti-de Sitter
spacetime, given by metric (\ref{(2.1)}) with conformal factor $({\cal K}X^0)^{-2}$ instead of $({\cal K}X^4)^{-2}$, then the induced
metric on a brane defined by $X^0=A(\eta),~X^i=x^i,~X^4=T(\eta)$ would have been (\ref{(2.5)}) with the condition (\ref{(2.3)}) replaced by
$T'=\sqrt{A'^2-1}\Longrightarrow A'^2\geq1\Longrightarrow H\geq {\cal K}$, a condition which is not fulfilled by
standard cosmological scenarios. Hence the bulk cannot be chosen to be a de Sitter spacetime, at least when the branes
are defined as above. A Minkowski bulk does not lead either to an acceptable cosmological scenario, see [3] and
e.g. [18].

Let us recall now how the extrinsic curvature of a brane is calculated. One introduces the normal $\bar n^A(x^\mu)$ to the
brane as
\be \bar g_{AB}|_{\bar\Sigma}\,\bar n^A\bar V^B_\mu=0\quad,\quad \bar g_{AB}|_{\bar\Sigma}\,\bar n^A\bar
n^B=1\quad\hbox{that is}\quad\bar n^A={\cal K}A\,(A',0,0,0,T')\lbl{(2.7)}\ee
where the sign has been chosen arbitrarily. The extrinsic curvature of the brane is
then defined as (introducing  $\bar D$, the covariant derivative associated with $\bar g_{AB}$) $\bar
K_{\mu\nu}=-\bar V^A_\mu \bar V^B_\nu \bar D_A\bar n_B$, that is, using the symmetry $\bar K_{\mu\nu}=\bar
K_{\nu\mu}$, 
\be\bar K_{\mu\nu}=-{1\over2}\left[\bar g_{AB}|_{\bar\Sigma}\,(\bar V^A_\mu \partial_\nu\bar n^B +\bar
V^A_\nu\partial_\mu \bar n^B) +\bar V^A_\nu\bar V^B_\mu\bar n^C(\partial_C\bar
g_{AB})|_{\bar\Sigma}\,\right]\lbl{(2.8)}\ee 
which gives
\bea\bar K_{\eta\eta}&=&{1\over {\cal K}A^2T'}(AA''-1-A'^2)=-{a^2\over\sqrt{{\cal K}^2+H^2}}\left({\cal
K}^2+{\ddot a\over a}\right)\nonumber\\
\bar K_{ij}&=&{T'\over {\cal K}A^2}\delta_{ij}=a^2\sqrt{{\cal K}^2+H^2}\,\delta_{ij}\,.\lbl{(2.9)}\eea

Physics, gravity and matter, are introduced in this up to now purely geometrical picture by means of Einstein's
equations, $G_{AB}=\kappa T_{AB}$, where $G_{AB}$ is Einstein's tensor, $\kappa$ a five dimensional gravitational
coupling constant and $T_{AB}$ the stress energy tensor of matter. The $AdS_5$ bulk is then
interpreted as a solution of Einstein's equations with matter a cosmological constant: $\kappa
T_{AB}|_{\rm bulk}=6{\cal K}^2\bar g_{AB}$. Furthermore, matter is introduced on the brane by means of the so-called $Z_2$
symmetry which amounts to 

(1) cut $AdS_5$ along the brane (see e.g. [10] or [17] for conformal diagrams and appendix A for embedding descriptions of
this cut), 

(2) keep the side between the brane and $X^4\to +\infty$ [10],

(3) make a copy of this ``half" $AdS_5$ spacetime and join it to the original along the brane (hence the description of
the bulk as a doubled-sided piece of $AdS_5$),

(4) integrate Einstein's equations across this singular surface and obtain the Lanczos-Darmois-Israel equations often
called ``junctions conditions" [19]  and thus get the stress energy tensor $\bar{\cal T}_{\mu\nu}$ of the matter on the
brane in terms of its extrinsic curvature as 
\be\kappa\left(\bar{\cal T}_\mu^\nu-{1\over3}\delta^\nu_\mu\,\bar{\cal T}\right)=2\bar K_\mu^\nu\lbl{(2.10)}\ee
where the
indices are raised by means of the inverse metric $a^{-2}\eta^{\mu\nu}$ and $\bar{\cal T}\equiv\bar{\cal
T}_\mu^\mu$. Condition (2) above together with the choice of sign in equation (\ref{(2.7)}) ensures that the energy density on the brane
is definite positive [20]. The spatial components of (\ref{(2.10)}) give, using (\ref{(2.8)}) and noting
$\bar{\cal T}^0_0=-\rho$, $\bar{\cal T}^i_j=p\delta^i_j$,
\be\kappa\rho =6{\cal K}\sqrt{1+A'^2}\quad\Longleftrightarrow\quad \kappa\rho =6\sqrt{{\cal K}^2+H^2}\,.\lbl{(2.11)}\ee
As for the $(0,\mu)$ components of (\ref{(2.10)}) they are equivalent to the conservation of $\bar{\cal T}_\mu^\nu$, because
matter in the bulk reduces to a cosmological constant (see e.g. [20]). They read
\be\dot\rho+3H(\rho+p)=0 \quad\Longleftrightarrow\quad\kappa p=2{{\cal K}\over T'}(AA''-3T'^2)\, .\lbl{(2.12)}\ee
One thus sees in particular that  Einstein's equations impose that minkowskian branes, such that $A(\eta)=$~constant,
must contain matter under the form of a ``tension" such that $\kappa\bar{\cal
T}^\mu_\nu=-6{\cal K}\delta^\mu_\nu$ [1]. More generally, an equation of state $\rho=6{\cal
K}/\kappa+\rho_m$,
$p=-6{\cal K}/\kappa+p_m$ with $p_m=p_m(\rho_m)$ being chosen, equations (\ref{(2.11)}-\ref{(2.12)}) together with (\ref{(2.3)}) give
$A(\eta)$ (or $a(t)$) as well as $\rho(\eta)$. Various cosmological scenarios can hence be built [2-5].

\section*{III Brane-bending in an anti-de Sitter bulk}
In this section we shall consider an unperturbed anti-de Sitter bulk that we shall describe using the conformally
minkowskian coordinates (\ref{(2.1)}). 
We therefore do not allow here for perturbations of the coordinate system in the bulk;
cf. section V for gauge related issues. On the other hand we allow for perturbations in the position of the brane - this is
the ``brane-bending" effect analyzed by e.g. [21] in the case of a minkowskian background brane. In other
words we consider in $AdS_5$ with metric (\ref{(2.1)}) a brane $\Sigma$ defined by
\be X^A=\bar X^A(x^\mu)+\epsilon^A(x^\mu)\lbl{(3.1)}\ee
where $\bar X^A(x^\mu)$ are given by (\ref{(2.2)}-\ref{(2.3)}) and where the
five ``small" functions
$\epsilon^A(x^\mu)$ can be conveniently decomposed along the four tangent vectors to the brane (\ref{(2.4)}) and its normal
(\ref{(2.7)}) according to
\be\epsilon^A=\xi^\lambda\,\bar V^A_\lambda+\zeta\,\bar n^A\lbl{(3.2)}\ee 
with $\xi^\lambda(x^\mu)$ and $\zeta(x^\mu)$ five arbitrary functions of the coordinates $x^\mu=(\eta,x^i)$
which parametrize the brane. Strictly speaking the tangent and normal vectors $\bar V^A_\lambda$ and $\bar n^A$ are
defined on the unperturbed brane only. The vectors which appear in (\ref{(3.2)}) are Lie transported to the perturbed brane.
(See [12] for equivalent parallel transport.)
  
When $\zeta=0$ there is no deformation of the brane and the perturbation $\epsilon^A$ amounts to a slight
change in its parametrization, which can be absorbed into the infinitesimal change of coordinates:
$x^\lambda=\tilde x^\lambda-\xi^\lambda$ (as it is easy to show explicitly; cf. appendix B). We shall therefore set $\xi^\lambda=0$ and
describe the deformation of the brane by the single function $\zeta$. Hence the gauge is completely fixed, in the brane
as well as in the bulk.

A short calculation then shows that the induced metric on the perturbed brane
\be ds^2|_4=\bar g_{AB}|_{\Sigma} \left[\bar V^A_\mu+\partial_\mu(\zeta \bar
n^A)\right]\left[\bar V^B_\nu+\partial_\nu(\zeta \bar n^B)\right] dx^\mu dx^\nu\lbl{(3.4)}\ee
can be expressed in terms of the  background brane  extrinsic curvature
$\bar K_{\mu\nu}$  (\ref{(2.8)}) as
\be ds^2|_4={1\over ({\cal K}A)^2}\left(\eta_{\mu\nu}+\gamma^{(p)}_{\mu\nu}\right)dx^\mu dx^\nu\quad\hbox{with}\quad
\gamma^{(p)}_{\mu\nu}=-2 ({\cal K}A)^2\zeta\bar K_{\mu\nu}
\lbl{(3.5)}\ee
where the index $(p)$ stands for perturbation of the {\it position} of the brane. (This perturbation cannot be gauged away,
unless the background brane extrinsic curvature tensor vanishes.)
 In terms of the scale factor, cosmic time and Hubble parameter it reads
\be ds^2|_4=-\left[1-{2\zeta\over\sqrt{{\cal K}^2+H^2}}\left({\cal K}^2+{\ddot a\over
a}\right)\right]dt^2+a^2\left(1-2\zeta\sqrt{{\cal K}^2+H^2}\right)\delta_{ij}dx^i dx^j.
\lbl{(3.6)}\ee
Hence, for a scale factor behaving as $t^p$, the induced metric remains bounded if the function $\zeta$ behaves as $t^q$,
$q\geq1$, at early times.

The normal vector to the perturbed brane and its intrinsic curvature $K^{(p)}_{ij}$ are obtained from the definitions
(\ref{(2.7)}-\ref{(2.8)}) with all bars dropped, apart from the one on $\bar g_{AB}$. One obtains, denoting 
the perturbation of the extrinsic curvature of the brane due to its bending as $\delta^{(p)}
K^i_j\equiv K^{i\,(p)}_j-\bar K^i_j$, 
\bea\delta^{(p)} K^i_j&=&({\cal K}A)^2\left[\partial^i_j\zeta+ \delta^i_j\left({A'\zeta'\over
A} +{A'^2\zeta\over A^2}\right)\right]\nonumber\\
&=&{1\over a^2}\partial^i_j\zeta+H(H\zeta-\dot\zeta)\,\delta^i_j.\lbl{(3.7)}\eea
(The indices of $K^{i\,(p)}_j$ are raised by means of the metric (\ref{(3.6)}) and $\partial^i_j\zeta\equiv\delta^{ik}\partial_{jk}\zeta$.)

\section*{IV Perturbing the geometry of the bulk}
In this section we consider a perturbed anti-de Sitter bulk with metric
\be ds^2|_5= g_{AB}dX^AdX^B={1\over ({\cal K}X^4)^2}(\eta_{AB}+h_{AB})dX^AdX^B\lbl{(4.1)}\ee
where, among the fifteen function $h_{AB}(X^C)$, five have been chosen to fix the gauge in the bulk and the
remaining ten are imposed not to be reducible to zero by a change of coordinates.  As for the brane
$\bar \Sigma$ it is defined by the {\it same} equations as in the unperturbed case, that is by equations 
(\ref{(2.2)}-\ref{(2.3)}). Of course
this brane, despite its notation,  is geometrically different from the unperturbed Robertson-Walker brane of section II.

The induced metric on $\bar \Sigma$ is
\be ds^2|_4=g_{AB}|_{\bar\Sigma}\,\bar V^A_\mu\bar V^B_\nu dx^\mu dx^\nu={1\over ({\cal K}A)^2}(\eta_{\mu\nu}+
\gamma^{(b)}_{\mu\nu})dx^\mu dx^\nu\,,\lbl{(4.2)}\ee
with 
\bea\gamma^{(b)}_{\eta\eta}&=&T'^2h_{00}|_{\bar\Sigma}+2T'A'h_{04}|_{\bar\Sigma}+A'^2h_{44}|_{\bar\Sigma}\nonumber\\
\gamma^{(b)}_{\eta i}&=&T'h_{0 i}|_{\bar\Sigma}+A'h_{4i}|_{\bar\Sigma}\nonumber\\
\gamma^{(b)}_{ij}&=&h_{ij}|_{\bar\Sigma}\,\lbl{(4.3)}\eea
where the index $(b)$ means that these perturbations are induced by the perturbations of geometry of the {\it bulk}.

As for the normal vector to the brane and its extrinsic curvature $K^{(b)}_{ij}$ they are again defined by 
(\ref{(2.7)}-\ref{(2.8)}) where,
here, all bars are kept, apart from the one on $g_{AB}$. One obtains for the perturbation of the extrinsic curvature
of the brane due to the perturbations of geometry of the bulk, $\delta^{(b)}K_{ij}\equiv K_{ij}^{(b)}-\bar K_{ij}$, 
\bea\delta K^{(b)}_{ij}&=&{1\over
2{\cal K}A}\left[A'(\partial_jh_{0i}+\partial_ih_{0j}-\partial_0h_{ij})|_{\bar\Sigma}\,
+T'(\partial_jh_{4i}+\partial_ih_{4j}-\partial_4h_{ij})|_{\bar\Sigma}\,\right]+\nonumber\\
&+&{1\over
{\cal K}A^2}\left\{\delta_{ij}\left[\,T'(A'^2-{\scriptstyle{1\over2}}T'^2)h_{44}|_{\bar\Sigma}+A'^3h_{04}|_{\bar\Sigma}+
{\scriptstyle{1\over2}}A'^2T'h_{00}|_{\bar\Sigma}\,\right]+T'h_{ij}\right\}\,.\lbl{(4.4)}\eea
This expression can be rewritten in a more compact and geometrical form as
\be\delta^{(b)}
K^i_j=\pi^i_j-{\scriptstyle{1\over2}} \sigma^i_j-a\left(H\gamma^{(b)}_\eta+{a\over2}
\sqrt{{\cal K}^2+H^2}\,\gamma^{(b)}
\right)\delta^i_j\lbl{(4.5)}\ee
where $\delta^{(b)}
K^i_j=K^{i\,(b)}_j-\bar K^i_j$ (indices being raised by means of the metric(\ref{(4.2)})\,) and where we have introduced
\be\sigma_{ij}=\bar n^A(\partial_Ah_{ij})|_{\bar\Sigma}\qquad\hbox{and}\qquad
\pi_{ij}={1\over2}[\bar n^A(\partial_jh_{Ai})|_{\bar\Sigma}+\bar n^A(\partial_ih_{Aj})|_{\bar\Sigma}]\,,\lbl{(4.6)}\ee
(with $\pi^i_j\equiv\delta^{ik}\pi_{jk}$,  $\sigma^i_j\equiv\delta^{ik}\sigma_{jk}$) as well as
\be\gamma^{(b)}_\eta=h_{AB}|_{\bar\Sigma}\,\bar n^A\bar V^B_\eta\qquad\hbox{and}\qquad 
\gamma^{(b)}=h_{AB}|_{\bar\Sigma}\,\bar n^A \bar n^B.\lbl{(4.7)}\ee
Using (\ref{(2.4)}) and (\ref{(2.7)}) one finds explicitly
\bea\gamma^{(b)}_\eta&=&{1\over {\cal K}^2a}\left[-H\sqrt{{\cal
K}^2+H^2}(h_{00}|_{\bar\Sigma}+h_{44}|_{\bar\Sigma}) +({\cal K}^2+2H^2)h_{04}|_{\bar\Sigma}\right]\nonumber\\
\gamma^{(b)}&=&{1\over {\cal K}^2a^2}\left[H^2h_{00}|_{\bar\Sigma}-2H\sqrt{{\cal K}^2+H^2}h_{04}|_{\bar\Sigma}
+({\cal K}^2+H^2)h_{44}|_{\bar\Sigma}\right]\,.\lbl{(4.8)}\eea

We postpone until section VII the interpretation of the perturbations of extrinsic curvature in terms of matter
perturbations on the brane.

\section*{V Gauge related issues}
In section III we considered a strictly anti-de Sitter bulk in conformally minkowskian coordinates and we perturbed the
position of the brane along its normal. In so doing, as we have seen, we fixed the gauge completely, in the bulk as well
as in the brane, and introduced a single function $\zeta(x^\mu)$ the effect of which on the induced metric and
extrinsic curvature of the brane cannot be gauged away and is given by equations (\ref{(3.5)}-\ref{(3.6)})
and (\ref{(3.7)}).

In the previous section we geometrically perturbed the bulk in a given coordinate system and fixed the position of the
brane in that system. We have thus introduced ten functions $h_{AB}(X^C)$ which characterize in a given coordinate
system a geometrical perturbation of the bulk, i.e. we imposed that they cannot be gauged away. Their effect on the
brane is given by equations (\ref{(4.2)}-\ref{(4.3)}) and (\ref{(4.4)}) or (\ref{(4.5)}). 

The brane bending perturbation adds a degree of freedom to these ten perturbations of the bulk 
since it can be present
in the geometrically unperturbed background $AdS_5$ bulk. We note in passing that this effect can be described
alternatively as a brane perturbation induced by a general coordinate shift performed in the background
$AdS_5$ bulk {\it without} changing accordingly the equation for the brane. See appendix B for details.

Hence, the eleven functions introduced describe completely the geometrical perturbations of the bulk and the position of the brane in that bulk. 
Now these eleven independent functions will be constrained in the next sections to satisfy Einstein's equations.
Imposing in a first step Einstein's equations in the five dimensional bulk, where matter is chosen to be a cosmological
constant, will reduce these eleven functions to six (according to the rule ``the gauge hits twice"). These six functions
will be interpreted as the five degrees of freedom of the $AdS_5$ gravitational waves plus a ``radion" describing the
motion of the brane. Imposing in a second step the $Z_2$ symmetry and Einstein's equations across the singular
brane will define the matter perturbations on the brane in terms of these six arbitrary functions, which is just the right
number to describe, in a given brane coordinate system, the most general four dimensional perturbed universe.

This counting can be generalized to any $N$-dimensional brane in a $D=N+1$-dimensional bulk. The
number of gauge independent metric perturbations (or, equivalently, the number of independent metric perturbations in a 
given gauge) in a $D$-dimen\-sional bulk is ${1\over2}D(D+1)-D={1\over2}D(D-1)$. The number of freely propagating
degrees of freedom (gravitational waves) in a $D$-dimensional bulk is ${1\over2}D(D-1)-D={1\over2}D(D-3)$. The
deformation of a $D-1=N$-dimensional brane is described by the normal vector $\zeta \bar n^A$, that is by one function.
Now we have ${1\over2}D(D-3)+1={1\over2}N(N-1)$ which is the number of gauge independent metric perturbations in a
$N$-dimensional brane.

We would like to argue at this point that normal gaussian coordinates (used by e.g. [7], [9], [12], [13]) in which the
perturbed bulk metric is written as
\be ds^2|_5=(g_{\mu\nu}+h_{\mu\nu})dx^\mu dx^\nu+dy^2\lbl{(5.1)}\ee
where $y=0$ is the position of the brane and where the explicit expression of the background anti-de Sitter metric
coefficients $g_{\mu\nu}$ can be found in [15], seem less appropriate to treat the problem at hand than the
conformally minkowskian coordinates advocated here. Indeed one can certainly use the form (\ref{(5.1)}) of the metric to study
linearized gravity on an anti-de Sitter background. However

(1) the linearized Einstein equations are much simpler, and their boundary conditions much easier to implement, when
written in conformally minkowskian coordinates (as recalled in the next section),

(2) imposing that the brane is at $y=0$ means choosing among all coordinate systems such that $h_{yy}=h_{y\mu}=0$,
the particular sub-class which is adapted to the bending of the brane [7]. This implies that when solving the
linearised Einstein equations in the bulk one can no longer simplify them by choosing the best adapted coordinate system
within the class (1) (like, for example, an harmonic system). In practice this means that one must solve the constraint equations in full generality. This
introduces an arbitrary function $\zeta(x^\mu)$ which encodes the brane bending effect as well as the transformation
to the coordinate system in which the solution of the constraint equations is simple and the brane located at
$y=\zeta(x^\mu)$.  (A similar procedure must be applied in synchronous gauge descriptions of the surface defining
reheating in inflationary scenarios, see e.g. [22] for further discussion of that point.)

When the background brane is Minkowski spacetime, as in the Randall-Sundrum scenario, conformally minkowskian and
gaussian normal coordinates are almost identical, so that objection (1) falls in that case. As for objection (2) it falls as
well since it is then as simple to solve the linearised Einstein equations in an harmonic gauge where the brane
is located at $y=\zeta(x^\mu)$ than in a gaussian normal gauge (cf. e.g. [21] and [23]). When, on the
other hand the background brane is a Robertson-Walker spacetime, the simplicity of the linearised Einstein equations in
conformally minkowskian coordinates discussed in section VI will, as we hope to convince the reader, compensate for the
slightly more complicated form of the junction conditions given in section VII.

\section*{VI Einstein's equations in the bulk}
The metric perturbations of the bulk we considered in section IV are now forced to obey Einstein's equations, matter
being chosen to be a cosmological constant: $G_{AB}=6{\cal K}^2g_{AB}$. The metric being given by (\ref{(4.1)}) their
linearization gives the following equations for the perturbations $h_{AB}$ everywhere outside the brane [23]
\bea{1\over2}& &\left[\partial_{AL}h^L_B+\partial_{BL}h^L_A-\partial_{AB}h-\Box_5 h_{AB}-
\eta_{AB}(\partial_{LM}h^{LM}-\Box_5 h)\right]\nonumber\\
-{6\over (X^4)^2}& &\eta_{AB}h_{44}
-{3\over2X^4}\left[\partial_Ah_{4B}+\partial_Bh_{4A}-\partial_4h_{AB}+
\eta_{AB}(\partial_4h-2\partial_Lh^L_4)\right]=0\lbl{(6.1)}\eea
where all indices are raised with $\eta^{AB}$, $h\equiv h^L_L$ and $\Box_5\equiv\partial_L\partial^L$.

These
equations must be solved in a given gauge. If we impose the conditions
\be h_{4A}=0\lbl{(6.2)}\ee
the fifteen functions $h_{AB}(X^A)$  reduce to the ten functions $h_{\mu\nu}(X^A)$ which satisfy
\bea\partial_{\rho\sigma}h^{\rho\sigma}-\Box_4h+{3\over X^4}\partial_4h&=&0\nonumber\\
\partial_4(\partial_\rho h^\rho_\mu-\partial_\mu h)&=&0\nonumber\\
\partial_{44}h-{1\over X^4}\partial_4h&=&0\nonumber\\
\Box_4h_{\mu\nu}+\partial_{44}h_{\mu\nu}-{3\over X^4}\partial_4h_{\mu\nu}&=&
\partial_{\mu\rho}h^\rho_\nu+\partial_{\nu\rho}h^\rho_\mu-\partial_{\mu\nu}h+
{\eta_{\mu\nu}\over X^4}\partial_4h\,.\lbl{(6.3)}\eea
The first three constraint equations are easily solved and one then chooses (a choice that one cannot make when using
gaussian normal coordinates) the coordinate system satisfying condition (\ref{(6.2)}) such that  the solution
reduces to
\be h\equiv\eta^{\rho\sigma}h_{\rho\sigma}=0\qquad,\qquad\partial_\rho h^\rho_\mu=0\,.\lbl{(6.4)}\ee
Hence the choice of gauge together with the constraint equations reduce the ten functions $h_{\mu\nu}(X^A)$ to five,
which represent the five degrees of freedom of $AdS_5$ gravitational waves.

As for the fourth evolution equation (\ref{(6.3)}) it is solved by separation of variables. Inserting the
ansatz
\be\hat h_{\mu\nu}=(mX^4)^2Z_2(mX^4)\,e_{\mu\nu}(k^i,m)\,e^{{\rm i}\,k_\rho X^\rho}\lbl{(6.5)}\ee
where  $m$ and $k^i$ are the four separation constants one obtains 
\be{d\over dX^4}{dZ_2\over dX^4}+{1\over X^4}{dZ_2\over dX^4}+\left(m^2-{4\over
(X^4)^2}\right)Z_2=0\lbl{(6.6)}\ee
 and
\be k^\rho k_\rho=-m^2\qquad\Longrightarrow\qquad k_0=-\sqrt{k_ik^i+m^2}\,.\lbl{(6.7)}\ee

For $m\neq 0$ the general solution of (\ref{(6.6)}), which represents the ``Kaluza-Klein excitations" introduced in this
context by [1], is a combination of Bessel functions of order 2 [24]
\be Z_2(mX^4)=a_m J_2(mX^4)+b_m N_2(mX^4)\lbl{(6.8)}\ee
where $a_m$ and $b_m$ are a priori arbitrary constants.

A word of caution is in order here. Since the conformally flat coordinates are
unsuited to describe the universal covering of $AdS_5$ (see appendix A and references quoted therein), one may have to
impose a boundary condition on $Z_2$ at
$X^4\to +\infty$. There does not seem to be an agreement on that point in the literature. For example the
authors of ref. [1], [21], [23] do not impose any condition at $X^4\to+\infty$.
As for [25] they impose  $b_m=0$ whereas [10], [26] or [27] 
choose $a_m=-{\rm i}b_m$ (cf. also [28]). We shall leave
this question open here and rather make the following remark. The ``zero-mode" perturbation is the $m=0$ bounded
and normalizable solution of (\ref{(6.6)}). It behaves as
$Z_2\propto (X^4)^{-2}$ so that in that case $\hat h_{\mu\nu}$ does not depend on  $X^4$ and can be considered
as the limit when $mX^4\to0$ of the Bessel function $N_2(mX^4)$. One may therefore advocate the condition
\be a_m=0\lbl{(6.9)}\ee
so that the bounded zero mode and the $m\neq0$ modes form a uniform family of states. 

Finally the constraint equations (\ref{(6.4)}) impose
\be k^\rho e_{\rho\mu}=0\qquad\hbox{and}\qquad
\eta^{\rho\sigma}e_{\rho\sigma}=0\,.\lbl{(6.10)}\ee

To summarize, the general solution of the linearized Einstein equations in an $AdS_5$ background is 
$$ ds^2|_5={1\over
({\cal K}X^4)^2}(\eta_{AB}+h_{AB})dX^AdX^B$$
with
\be h_{A4}=0\quad\hbox{and}\quad h_{\mu\nu}=\int\!\! dm \,d^3k\, \hat h_{\mu\nu}(X^A, k^i,m)\lbl{(6.11)}\ee 
where the mode, or gravitational wave, $\hat h_{\mu\nu}$ is given by  (\ref{(6.5)})  (\ref{(6.7)}) and (\ref{(6.8)}), 
where $e_{\mu\nu}$ is transverse and traceless (eq. (\ref{(6.10)})) and where the additional
condition  (\ref{(6.9)}) ensures that the massive modes tend to the bounded zero mode when $m\to0$. 

\section*{VII The Lanczos-Darmois-Israel equations}
We now turn to the matter perturbations on the brane. They are obtained by imposing the
Lanczos-Darmois-Israel equations (\ref{(2.10)}) with all bars dropped and with
$K^i_j=\bar K^i_j+\delta^{(p)}K^i_j+\delta^{(b)}K^i_j$, $\delta^{(p)}K^i_j$ and $\delta^{(b)}K^i_j$ being given 
respectively by equations (\ref{(3.7)}) and (\ref{(4.5)}), it being understood now that the perturbations in the bulk are given by
(\ref{(6.11)}). Hence, just using the fact that we chose the gauges $h_{4A}=0$, we have that
\bea{\kappa\over2}\delta\left({\cal T}^i_j-{1\over3}\delta^i_j{\cal T}\right)&=&
{1\over a^2}\,\partial^i_j\zeta+H\,\delta^i_j\left(H\zeta-\dot\zeta\right)+{H^2\over
2{\cal K}^2}\,\delta^i_j\,\sqrt{{\cal K}^2+H^2}\,h_{00}|_{\bar\Sigma}\,+\nonumber\\
& &{1\over
2{\cal K}a}\left[H(\partial_0h^i_j)|_{\bar\Sigma}-\sqrt{{\cal
K}^2+H^2}\,(\partial_4h^i_j)_{\bar\Sigma}-H(\partial_jh_0^i+\partial^ih_{0j})|_{\bar\Sigma}\right]\lbl{(7.1)}\eea
where, on the right-hand side, spatial indices are raised with $\delta^{ij}$. 

As for the $(0,\mu)$ components of the junction conditions, they are still equivalent to the conservation of
${\cal T}_{\mu\nu}$ because matter in the bulk reduces to a cosmological constant [20]
\be\nabla_\mu{\cal T}^\mu_\nu=0\lbl{(7.2)}\ee
$\nabla_\mu$ being the covariant derivative associated to the induced metric on the brane $a^2(\eta_{\mu\nu}+\gamma_{\mu\nu})$
with $\gamma_{\mu\nu}=\gamma_{\mu\nu}^{(p)}+\gamma_{\mu\nu}^{(b)}$.

Equations (\ref{(7.1)}-\ref{(7.2)}) are the central result of this paper. They look more complicated than analogous expressions obtained in
gaussian normal coordinates (cf. e.g. [7], [9], [11], [13]) but they include the brane bending effect explicitly and are
expressed in terms of the bulk gravitational waves written in conformally minkowskian coordinates which are known
and simple, as recalled in the previous section.

There are several ways to interpret these equations. If the gravitational waves in the bulk are given by some underlying
physics (they may be for example the zero point fluctuations of quantum gravitons) and if the perturbation of the
position of the brane is also governed by some theory then equations (\ref{(7.1)}-\ref{(7.2)}) just define a tensor which has no reason, a
priori, to be the stress-energy tensor of any realistic matter (although one can, of course, interpret it in terms of ``new
physics"). Conversely, if matter on the brane is imposed to be of a certain type, e.g. a scalar field or a perfect fluid with
or without topological defects etc., then equations (\ref{(7.1)}-\ref{(7.2)}) become ``junction conditions" which restrict the gravitational
waves in the bulk and the position of the brane to those which are compatible with the imposed brane stress-energy
tensor. Now it may be that some compromise has to be made for the junction conditions to have a solution. In fact this is
already the case when solving the background equations. As we saw in section II a Robertson-Walker brane can be
the edge of a given anti-de Sitter bulk, but at some price: matter on the brane has to include a fine-tuned, fairly
unphysical, tension in order for the scale factor of the brane to obey a reasonable quasi Friedmannian evolution equation.

To gain some insight on the restrictive aspect of equations (\ref{(7.1)}-\ref{(7.2)}) and in order to show how they can be solved 
explicitly, we shall for the sake of the example impose that matter in the brane reduces to a single scalar field
$\phi(x^\mu)$ with potential $V(\phi)$ plus a tension $\sigma$:
\be{\cal T}^\mu_\nu=\partial^\mu\phi\partial_\nu\phi-\delta^\mu_\nu
\left({1\over2}\partial_\rho\phi\partial^\rho\phi+V(\phi)\right)-\sigma\delta^\mu_\nu\,.\lbl{(7.3)}\ee
Setting $\phi(x^\mu)=\Phi(\eta)+\chi(x^\mu)$ one first obtains for the background brane equations (\ref{(2.11)}-\ref{(2.12)})
\bea\ddot\Phi+3H\dot\Phi+{dV\over d\Phi}&=&0\nonumber\\
{\kappa\over6}\left({\dot\Phi^2\over2}+V+\sigma\right)&=&\sqrt{{\cal K}^2+H^2}\lbl{(7.4)}\eea
in which, in order to recover standard cosmological scenarios, the tension must be fine tuned to $\kappa\sigma=6{\cal
K}$. A potential $V(\Phi)$ and initial conditions being chosen equations (\ref{(7.4)}) give $\Phi(t)$ and $a(t)$ (cf. e.g.
[29] where these equations are studied in detail).

At linear order in $\chi$ and the brane metric perturbations $\gamma_{\mu\nu}$ the left hand-side of equation (\ref{(7.1)}) reads
\be {\kappa\over2}\delta\left({\cal T}^i_j-{1\over3}\delta^i_j{\cal T}\right)  = {\kappa\over
6}\delta^i_j\left[\dot\Phi\dot\chi+\chi{dV\over d\Phi}+{\dot\Phi^2\over2}\gamma_{\eta\eta}\right]\,.
\lbl{(7.5)}\ee
Introducing the spatial tensor
\be F^i_j\equiv {1\over a^2}\,\partial^i_j\zeta+{1\over2{\cal K}a}\left[H(\partial_0h^i_j)|_{\bar\Sigma}-\sqrt{{\cal
K}^2+H^2}\,(\partial_4h^i_j)_{\bar\Sigma}-H(\partial_jh_0^i+\partial^ih_{0j})|_{\bar\Sigma}\right]\lbl{(7.6)}\ee
equation (\ref{(7.1)}) splits into a traceless and trace part
\bea F^i_j&=&{1\over3}\delta^i_j\, F\nonumber\\
F&=&{\kappa\over
2}\left[\dot\Phi\dot\chi+\chi{dV\over d\Phi}+{\dot\Phi^2\over2}\gamma_{\eta\eta}\right]
-3H(H\zeta-\dot\zeta)-{3H^2\over
2{\cal K}^2}\sqrt{{\cal K}^2+H^2}\,h_{00}|_{\bar\Sigma}\lbl{(7.7)}\eea
so that the junction conditions transform into (\ref{(7.6)}-\ref{(7.7)}) plus the conservation equations (\ref{(7.2)}), 
which is the Klein-Gordon equation for
$\phi$:
\be\ddot\chi-{1\over a^2}\Delta\chi+3H\dot\chi+{d^2V\over d\Phi^2}\chi+(\ddot\Phi+3H\dot\Phi)\gamma_{\eta\eta}
-{1\over a}\dot\Phi\partial_i\gamma^i_\eta+{\dot\Phi\over
2}(\dot\gamma_{\eta\eta}+\dot\gamma^i_i)=0\,.\lbl{(7.8)}\ee 

We now enter in equations (\ref{(7.6)}-\ref{(7.8)}) the explicit solution of the bulk Einstein equations. 
First, gathering (\ref{(3.5)}) and (\ref{(4.3)}), the brane metric perturbations are 
\bea\gamma_{\eta\eta}&=&{1\over{\cal K}^2}({\cal K}^2+H^2)h_{00}|_{\bar\Sigma}
+{2\zeta\over\sqrt{{\cal K}^2+H^2}}\left({\cal K}^2+{\ddot a\over a}\right)\nonumber\\
\gamma_\eta^i&=&{1\over {\cal K}}\sqrt{{\cal K}^2+H^2}h_0^i|_{\bar\Sigma}\nonumber\\
\gamma^i_j&=&h^i_j|_{\bar\Sigma}-{2\zeta}\,\sqrt{{\cal K}^2+H^2}\,\delta^i_j\,.\lbl{(7.9)}\eea
Second, the perturbations $h_{\mu\nu}$ are given by
(\ref{(6.11)}). More explicitly, we have for each mode  $\hat h_{\mu\nu}$
\bea\hat h_{\mu\nu}|_{\bar\Sigma}&=&\left({m\over{\cal K}a}\right)^2Z_2\left({m\over{\cal K}a}\right)
e_{\mu\nu}e^{{\rm i}\left(-\sqrt{k^2+m^2}\,T(t)+k_i x^i\right)}\nonumber\\
(\partial_0\hat h_{\mu\nu})|_{\bar\Sigma}&=&-{\rm i}\sqrt{k^2+m^2}\,\hat h_{\mu\nu}|_{\bar\Sigma}\nonumber\\
(\partial_i\hat h_{\mu\nu})|_{\bar\Sigma}&=&{\rm i}k_i\,\hat h_{\mu\nu}|_{\bar\Sigma}\nonumber\\
(\partial_4\hat h_{\mu\nu})|_{\bar\Sigma}&=&m\left({m\over{\cal K}a}\right)^2Z_1\left({m\over{\cal K}a}\right)
e_{\mu\nu}e^{{\rm i}\left(-\sqrt{k^2+m^2}\,T(t)+k_i x^i\right)}
\lbl{(7.10)}\eea
where $T(t)$ is given by (\ref{(2.3)}), that is $T(t)=\int\! dt\sqrt{{\cal K}^2+H^2}/{\cal K}a$, and where some
standard properties of the Bessel functions have been used [24].  Finally, we can without loss of generality consider only the
modes such that $k_1=k_2=0$, $k_3\equiv k$. The transverse and traceless properties of $e_{\mu\nu}$ then imply that
the five possible polarisations are characterised by $e_{11}$, $e_{12}$, $e_{13}$, $e_{23}$ and $e_{33}$, the other
components being $e_{0i}=-ke_{i3}/\sqrt{k^2+m^2}$, $e_{00}=k^2e_{33}/(k^2+m^2)$, and $e_{22}=-e_{11}-m^2e_{33}/(k^2+m^2)$.

We are now in a position to try and solve explicitly the junction conditions for each mode. 
The traceless equation (\ref{(7.7)}) first reduces for $m\neq0$ the five a priori possible polarisations to only one, characterized
by $e_{33}\equiv e(k,m)$. The others are
\bea e_{12}&=&e_{13}=e_{23}=e_{01}=e_{02}=0\nonumber\\
     e_{11}&=&e_{22}=-{1\over2}{m^2\over k^2+m^2}\,e\nonumber\\
     e_{03}&=&-{k\over\sqrt{k^2+m^2}}\,e\,\nonumber\\
     e_{00}&=&{k^2\over k^2+m^2}\,e\,.\lbl{(7.11)}\eea 
When $m=0$ on the other hand not only does $e_{33}$ survive but $e_{13}$ and $e_{23}$ as well. The latter two
polarisations will correspond to gravitational waves freely propagating in the brane (the so-called tensorial modes).

The traceless equation (\ref{(7.7)}) also forces the perturbation of the position of the brane to be a linear superposition of the
following modes
\be\zeta={e\over2{\cal K}k^2} \, a\left[{\rm i}{H\over\sqrt{k^2+m^2}}\left(k^2-{3\over2}m^2\right)\,G_2 
-m{\sqrt{{\cal K}^2+H^2}\over k^2+m^2}\left(k^2+{3\over2}m^2\right) \, G_1\right] \lbl{(7.12)}\ee 
where
\be G_{1,2}\equiv\left({m\over{\cal K} a}\right)^2Z_{1,2} \left({m\over{\cal K} a}\right) e^{{\rm
i}\left(-\sqrt{k^2+m^2}\,T(t)+k x^3\right)}\,.\lbl{(7.13)}\ee
The function $\zeta$ being now known, the induced metric $\gamma_{\mu\nu}$ on the brane, for $m\neq0$, is also
completely known in closed form via equations (\ref{(7.9)}-\ref{(7.13)}) in terms of two (or three) arbitrary functions $e(k,m)$,
$b_m(k,m)$ (and, should the occasion arise, $a_m(k,m)$). For the zero mode $m=0$, $e_{13}$, $ e_{23}$ and
$e_{33}\equiv e$ are arbitrary functions of $k$ and the above expression for
$\zeta$ becomes, since $a_m=0, b_m=1$ and $z^2N_2(z)\rightarrow -4/\pi$ as $z\to 0$,
\be\zeta=-{2{\rm i}e\over\pi k{\cal K}}\, Ha\, e^{-{\rm i}k(T-x^3)}\,.\lbl{(7.14)}\ee

What remains to be determined is the scalar field perturbation $\chi$ which must be extracted from the trace equation
(\ref{(7.7)}) and the Klein-Gordon equation (\ref{(7.8)}). One can proceed as follows. Inserting the expressions for $\zeta$
and the induced metric obtained above one can write (\ref{(7.7)}-\ref{(7.8)}) as
\be\dot\Phi\dot\chi+\chi {dV\over d\Phi}=F_1(t)\quad,\quad\ddot\chi+3H\dot\chi+\chi\left({k^2\over
a^2}+{d^2V\over d\Phi^2}\right)=F_2(t)\lbl{(7.15)}\ee
where $F_1$ and $F_2$ are known function of $t$. Hence
\be\chi={1\over6H{dV\over d\Phi}-\dot\Phi{k^2\over a^2}}(\dot F_1+6HF_1-\dot\Phi F_2)\,.\lbl{(7.16)}\ee
In the case of the zero mode $m=0$ one obtains
\be\chi={2{\rm i}e\over\pi k{\cal K}}\, a\dot\Phi\sqrt{{\cal K}^2+H^2}\, e^{-{\rm i}k(T-x^3)}\lbl{(7.17)}\ee
and it can be checked that the expression found is indeed a solution of (\ref{(7.15)}). For massive modes $m\neq0$ the algebra is more
involved. Since our purpose in this paper was just to write the Lanczos-Darmois-Israel equations in such a way as to be
able to try and solve them we shall present elsewhere the $m\neq0$ case as well as a comparison with the results of
ordinary chaotic inflation. 

% ---- End of paper -------------------------------------------------------------------------------------------------------

\section*{Acknowledgements}
 We warmly thank David Langlois for arousing our interest in brane cosmologies and
discussing with us the subtleties of gaussian normal coordinates, as well as Thibaut Damour for comments on the
restrictive aspect of the junction conditions. We also acknowledge discussions with the participants of the UK Cosmology
meeting held in Durham in September 2000, in particular Anne Davis, Jaume Garriga, Ruth Gregory and David Wands.

% ---- Appendices  -------------------------------------------------------------------------------------------------------

\appendix

\section{}
It can be useful, in particular when considering global properties of the brane or boundary conditions on the bulk
perturbations, to embed $AdS_5$ in a higher, six dimensional, flat space. It is known [31] that the surface defined
by 
\be (y^0)^2+\delta_{ij}y^iy^j-(y^4)^2-z^2=-{1\over {\cal K}^2}\lbl{(A.1)}\ee
in the six dimensional flat space with metric
\be ds^2|_6=(dy^0)^2+\delta_{ij}dy^idy^j-(dy^4)^2-dz^2\lbl{(A.2)}\ee
is $AdS_5$. This space contains closed timelike curves: the circles $(y^4)^2+z^2=$~constant. One goes round this difficulty
by introducing an integer ``winding number" which increases by 1 each time one goes round the circle. One thus
obtains the $AdS_5$  universal covering space [26].

The intersections of the planes $y^0=$~constant (or $y^i=$~constant) with the surface (\ref{(A.1)}) are four dimensional simply
connected hyperboloids of smallest radius $\sqrt{{\cal K}^{-2}+(y^0)^2}$. The sections $y^4=$~constant (or $z=$~constant) are
either four dimensional simply connected hyperboloids, or cones, or else doubly connected hyperboloids, depending on
whether $(y^4{\cal K})^2<1$, $(y^4{\cal K})^2=1$ or $(y^4{\cal K})^2>1$. 

If one parametrises the surface (\ref{(A.1)}) by the coordinates $X^A$ such that [25], [26]
\bea y^0&=&{1\over2X^4}\left[{\cal K}^{-2}+{(X^0)^2-\delta_{ij}X^iX^j-(X^4)^2}\right]\nonumber\\
y^i&=&{X^i\over {\cal K}X^4}\nonumber\\ 
y^4&=&{X^0\over {\cal K}X^4}\nonumber\\
z&=&{1\over2X^4}\left[{\cal K}^{-2}-{(X^0)^2+\delta_{ij}X^iX^j+(X^4)^2}\right]\lbl{(A.3)}\eea
its induced metric is conformally minkowskian:
\be ds^2|_5={1\over ({\cal K}X^4)^2}\,\eta_{AB}\,dX^AdX^B.\lbl{(A.4)}\ee
Note that the  plane $X^4=\infty\Leftrightarrow y^0+z=0$ is a coordinate singularity.

A de Sitter brane (such that $A(\eta)=\eta$, see equations (\ref{(2.2)}-\ref{(2.5)})\,)  is the intersection of the surface 
(\ref{(A.1)}) with the plane
$y^4=-\sqrt2/{\cal K}$ which is the familiar four dimensional simply connected hyperboloid of smallest radius $1/{\cal
K}$  embedded in a five dimensional Minkowski flat space [32]. The Minkowski brane
$A(\eta)=1/{\cal K}$ is the intersection of the surface (\ref{(A.1)}) with the plane $y^0+z=1/{\cal K}$. The Randall-Sundrum
[1] spacetime is obtained by keeping only the region between the brane and the coordinate singularity 
$X^4=+\infty$ [10].

If one now parametrises the surface (\ref{(A.1)}) by the coordinates $(\tau,r,\chi,\theta,\phi)$ such that
\bea y^0&=&r\sin\chi\sin\theta\sin\phi\quad ; \quad y^1=r\sin\chi\sin\theta\cos\phi\nonumber\\
y^2&=&r\sin\chi\cos\theta\quad\qquad ; \quad y^3=r\cos\chi \nonumber\\
y^4&=&\sqrt{1+r^2}\sin\tau\quad\quad~ ; \quad z=\sqrt{1+r^2}\cos\tau\lbl{(A.5)}\eea
the induced metric  is Schwarzschild-like
\be ds^2|_5=-(1+r^2)d\tau^2+{dr^2\over 1+r^2}+r^2(d\chi^2+\sin^2\chi d\theta^2+\sin^2\chi\sin^2\theta
d\phi^2).\lbl{(A.6)}\ee
By letting the coordinate $\tau$ vary from $-\infty$ to $+\infty$ one covers the $AdS_5$ universal covering space
without having to introduce the winding number.

The Schwarschild-like coordinates $(\tau,r,\chi,\theta,\phi)$ are therefore best suited to study the asymptotic
properties of quantum fields or classical gravitational waves in $AdS_5$ [33] and care must
be exercised when one uses the technically simpler conformally minkowskian coordinates $X^A$. 
The Schwarzschild-like coordinates are also well suited to the study of Robertson-Walker branes
with {\it closed}  spatial sections (they are simply defined by $r=a(\eta),~\tau=t(\eta)$ with $t(\eta)$ chosen so that
$\eta$ is conformal time [16]).

On the other hand the conformally minkowskian coordinates $X^A$ are better suited to the study, to which we confine
ourselves here, of  Robertson-Walker branes with {\it flat} spatial sections as well as that of the Randall-Sundrum
minkowskian brane.

\section{}
We show here explicitly the effect of a coordinate
change in the bulk on the induced metric of the brane and on its extrinsic curvature.

Let us consider the infinitesimal change of coordinates in the bulk $\tilde X^A\to X^A=\tilde X^A-\epsilon^A$,
$\epsilon^A(X^C)$ being five arbitrary functions of the coordinates $X^C$, {\it without} changing accordingly the equation
for the brane that we still define as in section IV by
$X^A=\bar X^A(x^\mu)$. Then the induced perturbation of the bulk metric is just the Lie derivative 
\be h_{AB}=-2\eta_{AB}\,{\epsilon^4\over
X^4}+\eta_{AC}\partial_B\epsilon^C+\eta_{BC}\partial_A\epsilon^C.\lbl{(B.1)}\ee 
The corresponding change of the induced
metric of the brane is obtained from equation (\ref{(4.3)}) and reads
\bea\gamma^{(b)}_{\eta\eta}&=&{2\epsilon^4|_{\bar\Sigma}\,\over
A}-2T'\partial_\eta\epsilon^0|_{\bar\Sigma}\, +2A'\partial_\eta\epsilon^4|_{\bar\Sigma}\,\nonumber\\
\gamma^{(b)}_{\eta
i}&=&-T'\partial_i\epsilon^0|_{\bar\Sigma}\,+A'\partial_i\epsilon^4|_{\bar\Sigma}\,+\delta_{ij}\partial_\eta\epsilon^j|_{\bar\Sigma}\,\nonumber\\
\gamma^{(b)}_{ij}&=&-2\delta_{ij}\,{\epsilon^4|_{\bar\Sigma}\,\over
A}+\delta_{jk}\partial_i\epsilon^k|_{\bar\Sigma}\,+\delta_{ik}\partial_j\epsilon^k|_{\bar\Sigma}\,.\lbl{(B.2)}\eea
As for the change of its extrinsic curvature it is given by (\ref{(4.4)}) and reads
\bea\delta^{(b)}K_{ij}&=&-A'\partial_{ij}\epsilon^0|_{\bar\Sigma}\,+T'\partial_{ij}\epsilon^4|_{\bar\Sigma}\,
+{T'\over
A}\left(\delta_{jk}\partial_i\epsilon^k|_{\bar\Sigma}\,+\delta_{ik}\partial_j\epsilon^k|_{\bar\Sigma}\,\right)+\nonumber\\
&+&{\delta_{ij}\over A}\left(A'T'\partial_\eta\epsilon^4|_{\bar\Sigma}\,-A'^2\partial_\eta\epsilon^0|_{\bar\Sigma}\,-
{2T'\over A}\epsilon^4|_{\bar\Sigma}\,\right)\,.\lbl{(B.3)}\eea
We can now decompose, as we did in section III,  $\epsilon^A|_{\bar\Sigma}$ along the tangent and normal vectors to the
brane as
\be\epsilon^A|_{\bar\Sigma}=\xi^\lambda\bar V^A_\lambda+\zeta \bar n^A.\lbl{(B.4)}\ee
It is then easy to see that if  $\zeta(x^\mu)=0$, so that $\epsilon^A|_{\bar\Sigma}=(T'\xi^\eta,\xi^i,A'\xi^\eta)$,
then $\gamma^{(b)}_{\mu\nu}$ and $\delta^{(b)}K_{ij}$ as given by (\ref{(B.2)}) and (\ref{(B.3)}) 
are just the Lie derivatives of the brane metric and its extrinsic curvature with respect to the vector field $\xi^{\mu}$,
and therefore describe the change in the components of these tensors under the coordinate shift
$\tilde x^\mu\to x^\mu=\tilde x^\mu-\xi^\mu$ on the brane. 
This result is geometrically obvious. Indeed $\zeta=0$ means
that the coordinate change in the bulk is such that the grid is moved parallely to the surface  $X^A=\bar
X^A(x^\mu)$ which defines the brane. Hence the brane is geometrically unperturbed by this operation.

On the other hand if $\xi^\lambda=0$, so that $\epsilon^A|_{\bar\Sigma}={\cal K}A(A'\zeta,0,0,0,T'\zeta)$, the
expressions (\ref{(B.2)}) and (\ref{(B.3)}) for $\gamma^{(b)}_{\mu\nu}$ and $\delta^{(b)}K_{ij}$  reduce to 
\bea\gamma^{(b)}_{\mu\nu}&=&-2({\cal K}A)^2\zeta\bar K_{\mu\nu}\nonumber\\
\delta^{(b)}K_{ij}&=&\partial_{ij}\zeta+\delta_{ij}\left({A'\zeta'\over
A} -{2\zeta\over A^2}-{A'^2\zeta\over A^2}\right).
\lbl{(B.5)}\eea
As expected they are identical to the perturbations due to a change of the position of the brane studied in section III
and given by equations (\ref{(3.5)}) and (\ref{(3.7)}). (For related views on the relationship between gauge and brane bending
effects, see e.g. [21], [27], [23], [28].)

% ---- Bibliography -------------------------------------------------------------------------------------------------------

\end{document}